\documentclass[aps,prl,preprint,nopacs,superscriptaddress]{revtex4}
\usepackage{graphicx}
\usepackage{verbatim}
\usepackage{mathrsfs}
\usepackage{amsmath,amsfonts,amssymb}
\usepackage{graphicx}

\def\3{2.8in}    
\def\2{2.5in}
\def\4{3.0in}

\def \beq {\begin{equation}}
\def \eeq {\end{equation}}
\begin{document}

\title{Atomic-scale visualization of quasiparticle interference on a type-II  Weyl semimetal surface}

\author{Hao~Zheng\footnote{These authors contributed equally to this work.}}
\affiliation {Laboratory for Topological Quantum Matter and Spectroscopy (B7), Department of Physics, Princeton University, Princeton, New Jersey 08544, USA}

\author{Guang~Bian $^*$}
\affiliation {Laboratory for Topological Quantum Matter and Spectroscopy (B7), Department of Physics, Princeton University, Princeton, New Jersey 08544, USA}

\author{Guoqing~Chang $^*$}
\affiliation {Centre for Advanced 2D Materials and Graphene Research Centre National University of Singapore, 6 Science Drive 2, Singapore 117546}
\affiliation {Department of Physics, National University of Singapore, 2 Science Drive 3, Singapore 117542}

\author{Hong~Lu}
\affiliation {International Center for Quantum Materials, School of Physics, Peking University, China}

\author{Su-Yang~Xu}
\affiliation {Laboratory for Topological Quantum Matter and Spectroscopy (B7), Department of Physics, Princeton University, Princeton, New Jersey 08544, USA}

\author{Guangqiang~Wang}
\affiliation {International Center for Quantum Materials, School of Physics, Peking University, China}

\author{Tay-Rong~Chang}
\affiliation{Department of Physics, National Tsing Hua University, Hsinchu 30013, Taiwan}

\author{Songtian~Zhang}
\affiliation {Laboratory for Topological Quantum Matter and Spectroscopy (B7), Department of Physics, Princeton University, Princeton, New Jersey 08544, USA}

\author{Ilya~Belopolski}
\affiliation {Laboratory for Topological Quantum Matter and Spectroscopy (B7), Department of Physics, Princeton University, Princeton, New Jersey 08544, USA}

\author{Nasser~Alidoust}
\affiliation {Laboratory for Topological Quantum Matter and Spectroscopy (B7), Department of Physics, Princeton University, Princeton, New Jersey 08544, USA}

\author{Daniel~S.~Sanchez }
\affiliation {Laboratory for Topological Quantum Matter and Spectroscopy (B7), Department of Physics, Princeton University, Princeton, New Jersey 08544, USA}

\author{Fengqi~Song}
\affiliation{National Laboratory of  Solid State microstructures, Collaborative Innovation Center of Advanced Microstructures and Department of Physics, Nanjing University, Nanjing, 210093, China}

\author{Horng-Tay~Jeng}
\affiliation{Department of Physics, National Tsing Hua University, Hsinchu 30013, Taiwan}
\affiliation{Institute of Physics, Academia Sinica, Taipei 11529, Taiwan}

\author{Nan~Yao}
\affiliation {Princeton Institute for the Science and Technology of Materials, Princeton University, 70 Prospect Avenue, Princeton, New Jersey 08540, USA} 

\author{Arun~Bansil}
\affiliation{Department of Physics, Northeastern University, Boston, Massachusetts 02115, USA}

\author{Shuang~Jia}
\affiliation {International Center for Quantum Materials, School of Physics, Peking University, China}
\affiliation {Collaborative Innovation Center of Quantum Matter, Beijing,100871, China}

\author{Hsin~Lin}
\affiliation {Centre for Advanced 2D Materials and Graphene Research Centre National University of Singapore, 6 Science Drive 2, Singapore 117546}
\affiliation {Department of Physics, National University of Singapore, 2 Science Drive 3, Singapore 117542}

\author{M. Zahid~Hasan}
\email{mzhasan@princeton.edu}
\affiliation {Laboratory for Topological Quantum Matter and Spectroscopy (B7), Department of Physics, Princeton University, Princeton, New Jersey 08544, USA}

\begin{abstract}
We combine quasiparticle interference simulation (theory) and atomic resolution scanning tunneling spectro-microscopy (experiment) to visualize the interference patterns on a type-II Weyl semimetal Mo$_{x}$W$_{1-x}$Te$_2$ for the first time. 
Our simulation based on first-principles band topology theoretically reveals the surface electron scattering behavior.  
We identify the topological Fermi arc states and reveal the scattering properties of the surface states in Mo$_{0.66}$W$_{0.34}$Te$_2$. 
In addition, our result reveals an experimental signature of the topology via the interconnectivity of bulk and surface states, which is essential for understanding the unusual nature of this material.

\end{abstract}

\maketitle
Recent discovery of type-I Weyl fermions in the TaAs class of materials has generated a flurry of new research directions \cite{Weyl1, Weyl2, Weyl3, Weyl4, Weyl5, rev1, rev2, TaAs1, TaAs2, ARPES-TaAs1, ARPES-TaAs2}. Many important predictions including Weyl cone,  Fermi arc,  chiral anomaly effect, and novel quasiparticle interference (QPI) were experimentally observed \cite{ARPES-TaAs1, ARPES-TaAs2, CA1, CA2, STM-NbP, STM-TaAs1, STM-TaAs2, STM-TaAs3}. Very recently, a new type of Weyl quasiparticle was predicted in WTe$_2$, MoTe$_2$ and their alloys \cite{WTe2, MoTe1, MoTe2, MoTe3}. The novelty is that this type (type-II) of Weyl fermions breaks Lorentz symmetry, and thus can not exist as a fundamental particle in nature. Such an excitation can emerge in a crystal as low-energy quasiparticles. Theory predicts that type-II Weyl semimetals host a number of unusual effects, \textit{e.g.} a new type of chiral anomaly, unconventional anomalous Hall effect and  interaction-induced emergent Lorentz invariant properties, which are not possible in type-I Weyl semimetals \cite{typeII1, typeII2, typeII3}. Thus the experimental investigation of the MoWTe-class of materials is desirable.

Scanning tunneling microscopy/spectroscopy (STM/S) is a vital tool for the investigation and illumination of various key properties  of a topological matter \cite{STM1, STM2, STM3, STM4, STM5}. The Fermi arc surface state, which is the topological fingerprint of Weyl semimetals, is predicted to exhibit exotic interference behavior in tunneling spectroscopy and magneto-transport measurements \cite{arc1, arc2, QPI1, QPI2}. Another unique property of a Weyl semimetal is the topological connection. An electron in a Fermi arc surface state, when moves to (or is scattered to) the Weyl node, will sink into the bulk and travel to the opposite surface \cite{arc1}. These features are interesting in connection to their QPI. 

We employed low temperature STM/S to investigate the QPI patterns in Mo$_x$W$_{1-x}$Te$_2$. The Fermi arc-derived quantum interference patterns are identified. We also performed comprehensive first-principle band structure calculations and QPI simulations on this material for the first time. Combination of our experimental and theoretical results reveals signatures of the predicted unique topological connection in this material. 

Single-crystalline Mo$_{0.66}$W$_{0.34}$Te$_2$ samples were grown by chemical vapor transport method. After being cleaved at 79~K, they were transferred \textit{in vacuo} to STM (Unisoku) at 4.6~K. dI/dV signals were acquired through a lock-in technique with a modulation at 5~mV and 1~kHz. Experimental QPI maps were generated by symmetrizing the Fourier transformed dI/dV maps (Fig. S1). First-principle based tight binding model simulations were used to obtain the electronic band structure. The alloy Mo$_{x}$W$_{1-x}$Te$_2$ was calculated by interpolation of the tight-binding model matrix elements of WTe$_2$ and MoTe$_2$. The theoretical QPI patterns are the restricted joint density of states, which removes all of the spin-flipping scattering vectors (see more details in the supplementary information).

Figs. 1(a) and 1(b) present the typical morphology of the cleaved Mo$_{0.66}$W$_{0.34}$Te$_2$(001) surface. There are only four point defects observed on the atomically ordered lattice, which confirms the high quality of our samples. In vicinity to the defect (Figs. 1 (c) and (d)), we observe a butterfly-like protrusion in the empty state image, and a depression area, which breaks the atomic row, in the occupied state. Since Mo$_x$W$_{1-x}$Te$_2$ is naturally cleaved at a Te-terminated surface, the point defect is attributed to a Te-vacancy. From the high resolution STM image of the occupied state (Fig. 1(e)), which probes the surface anions, we are able to clearly resolve an array of alternating atomic rows of extended (bright) and localized (dimmer) wave functions.The measured lattice constants (a=0.35~nm, b=0.63~nm) are consistent with the Te terminated T$_d$ phase of Mo$_x$W$_{1-x}$Te$_2$ \cite{lattice}. The simulated STM image in Fig. 1(f) reproduces the surface structure of the alternating Te-atom rows. We also calculate the density of state, considering only the top Te layer. A typical dI/dV spectrum in Fig. 1(h) displays finite conductance at zero bias, which indicates the (semi-)metallic behavior of the samples. The measured data agrees qualitatively with the simulation, which confirms structural properties of our Mo$_x$W$_{1-x}$Te$_2$ sample, a candidate for the type-II Weyl semimetal.

As shown in Fig. 2, the type-I Weyl cone consists of well separated upper and lower branches, and the constant energy contour (CEC) at the energy of Weyl node is a single point. By contrast, the type II cone is a heavily tilted in k-space, leading to the existence of projected bulk pockets (right and left branches of the Weyl cone) on the CEC at the Weyl node energy (and in a large energy range). Mo$_x$W$_{1-x}$Te$_2$ is predicted to be a type-II Weyl semimetal \cite{MoTe2}. Here, we focus on a x=0.66 sample. We uncover eight Weyl nodes in total. Four are located at 15~meV above the Fermi level (W$_1$), while the other four nodes sit at 62~meV (W$_2$).  On the CEC at 15~meV (Fig. 2(c)), we find a typical type-II Weyl semimetal feature, the coexistence of projected bulk states and surface states. According to the penetration depth (see more details in Fig. S2), we are able to identify the two bright yellow semi-circular contours in Fig. 2 (c) as surface states and the remaining light-blue pockets (one bowtie-shaped hole pocket and two elliptical electron pockets) in Fig. 2 (c) as projected bulk states. We enlarge the surface state in the vicinity of the two W$_1$ nodes in Figs. 2(d) and S2, and find that the surface band contour is split into three segments with tiny gaps in between. The middle segment behaves as a single curve connecting one pair of W$_1$ nodes and is therefore identified as the Fermi arc. In Fig. 2 (e), we plot the edges of the projected bulk bands (the branches of the type-II Weyl cone), and clearly demonstrate they touch each other at the Weyl nodes. In addition, from the energy-momentum dispersion in Fig. 2(f), one can see the tilted cone in the band structure. Taking these evidences together, we theoretically establish the type-II Weyl state in our Mo$_{0.66}$W$_{0.34}$Te$_2$ compound.

We perform dI/dV mapping on the Mo$_{0.66}$W$_{0.34}$Te$_2$(001) surface at various voltages and Fourier transform these maps to gain insight of the QPI information in Figure 3. Figs. 4(a)-(c) exhibit the experimental QPI maps acquired at 50~meV, 100~meV and 200~meV above Fermi level. The patterns in the red rectangles arise from the intra-first BZ quasiparticle scatterings, while patterns close to the Bragg points (Q$_x$,Q$_y$)=(0,$\pm\frac{2\pi}{b}$) with weaker intensities arise from the inter-BZ scattering and are replicas of the central features (Fig. S4). For simplicity, we restrict our discussion to the intra-BZ scattering (inside the rectangles) in the rest of this paper. At all energies (Fig. 3), the experimental QPIs  show simple and clean patterns, in contrast to TaAs \cite{STM-NbP, STM-TaAs1, STM-TaAs2, STM-TaAs3}. Specifically, all images consist of only three main pockets: one elliptical pocket in the center and two crescent-shaped contours located on the left and right sides of the central ellipse. The diameters of the crescents increase with bias voltage, which proves the electron like (instead of hole like) surface state. We perform model calculations to obtain the theoretical QPI patterns. Figs. 4(d)-(f) produce exactly the same number (three) of QPI pockets at the same locations in Q-space, and thus agree well with the experiments. Additionally, the calculated QPI at 100~meV (Fig.  4(e)) remarkably reproduces all dominant features in the measurement (Fig.  4(b)), namely, the central ellipse and the two side crescents. Moreover, the crescent displays a ``3"-shape rather than a ``$)$" shape,  the weak central feature as marked by the red arrow  is also reproduced in the simulation in Fig.  4(e). To further examine the evolution of the QPI features, we study the energy-scattering vector (E-Q) dispersion. In Fig.  4(g) the data in the region between -$\frac{\pi}{b}$ and $\frac{\pi}{b}$ corresponds to the intra-BZ scattering. In this region, the QPI signal displays as a V-shaped dispersion, with the vertex of the V located at Q=0. The linear edges of the V-feature marked by white arrows refer to the two cutting points on the crescent pocket, which are indicated by white arrows in Fig.  4(b). Besides this strongly dispersed QPI feature, we also reveal an additional weakly dispersed feature, which is denoted by the red arrow and corresponds to the pocket indicated by the red arrow in Fig.  4(b). The calculated E-Q dispersion reproduces both features in the experimental data in a wide energy range.

Fig. 5 shows the QPI map at 50~meV, which is between the energies of W$_1$ and W$_2$. At this energy, the Fermi arcs give clear interference signals. In a type-II Weyl semimetal, the large projected bulk pockets always appear in the surface CEC. In the calculated ``complete" CEC at 50~meV of the Mo$_{0.66}$W$_{0.34}$Te$_2$(001) surface, which includes both bulk and surface states as shown in Fig. 5(a),  we plot two dominant scattering vectors, namely $\vec{Q_1}$ connecting two electron pockets of the projected Weyl cone, and $\vec{Q_2}$ connecting the Weyl electron branch and the topologically trivial pocket at the $\bar{Y}$ points. In the simulated QPI pattern (Fig. 5(b)), one can clearly distinguish the $\vec{Q_1}$- and $\vec{Q_2}$-induced features. But the simulated QPI map consists of seven pockets,  more than what was observed in experiment (three pockets). Hence, it can not be a correct interpenetration of our observation. However, the situation is improved by removing the bulk bands and taking only the surface states into account (Fig. 5 (c)). The surface CEC is comprised only of two large semicircular-shaped contours and four small arc-like pockets. The dominant scattering vectors are $\vec{Q_3}$, which represents the scattering between two Fermi arc derived surface contours, and $\vec{Q_4}$, which links the topological surface state to the trivial state. Both scattering processes involve the electrons in the Fermi arc. Therefore, the $\vec{Q_3}$- and $\vec{Q_4}$-derived QPI features inside the white rectangles in Fig. 5(d) serve as an explicit evidence of the QPI signal from Fermi arcs. In the QPI data in Fig. 5(e), we indeed observe these features. More importantly, the white rectangles in Figs. 4(d) and 4(e) are located in same positions and are the same size. This proves that we have observed QPI pockets with comparable dimensions appearing at the predicted locations, providing more solid evidence of the detection of Fermi arc in our experiment. The subtle differences in the fine QPI features may originate from the commonly used simple assumptions in the calculation \cite{ref}. We emphasize that because Mo$_{x}$W$_{1-x}$Te$_2$ has a simple  band structure, the major feature of the surface-state-derived QPI pattern is robust. In addition, the agreement between our theory and experiment is remarkably good compared to previous QPI results on other materials. 

Furthermore, comparison of the two simulations in Figs. 5(b) and (d) to the experimental data (Fig. 5(e)) suggests that the contribution of the bulk states to the QPI signal is negligible. This can be attributed to the difficulty of establishing an interference between a three-dimensional bulk electronic wave and a two-dimensional surface wave. In other words, when a surface electron, which initially occupies a state in the Fermi arc, is scattered by a point defect into a bulk state, it loses its surface character and diffuses into the bulk. In this sense, the bulk Weyl pocket behaves like a sink of surface electrons, which is a signature of the topological connection between a Weyl cone and Fermi arc \cite{arc1, STM-TaAs1, STM-TaAs2}. On a type-I Weyl semimetal, the CEC at the energy of Weyl point consists of a point like bulk state, therefore the sinking effect is not prominent. By contrast, the type-II Weyl cones are heavily tilted, which gives rise to large areas of projected bulk pockets, and thus, this phenomenon should be more pronounced on a type-II Weyl semimetal surface. 

In summary, we present theoretical QPI simulations and STM results on Mo$_{x}$W$_{1-x}$Te$_2$ illustrating its complex electronic structure for the first time, which are complementary to the recent ARPES measurements \cite{MoTe2-ARPES1, MoTe2-ARPES2}.
Our QPI measurements directly discern the topological Fermi arcs. 
Taken together, our calculations and experiment data suggest that the interference pattern is dominated by surface states, whereas the contribution from bulk states to QPI is negligible, indicating the topological connection between the Weyl bulk states and Fermi arc surface states. 
Our results on Mo$_{x}$W$_{1-x}$Te$_2$ establish a platform for further study of novel spectroscopic, optical, and transport phenomena that emerge in this compound. 

After the completion of this theoretical plus experimental STM paper, we became aware of partial experimental STM data in a concomitant mainly-ARPES paper \cite{MoTe2-ARPES3}.

H.Z., G.B., S.-Y.X., I.B., N.A., D.S.S., S.Z. and M.Z.H. are supported by the Gordon and Betty Moore Foundations EPiQS Initiative through Grant GBMF4547 and U.S. National Science Foundation (NSF) Grant No. NSFDMR-1006492. H.Z also thanks the support from National Natural Science Foundation of China (No.11674226). G.C., and H.L. thank the National Research Foundation, Prime Ministers Office, Singapore, under its NRF fellowship (NRF Award No. NRF-NRFF2013-03). H.L., G.W., and S.J. are supported by National Basic Research Program of China (Grant No. 2013CB921901 and 2014CB239302). N.Y. is supported in part by funding from the Princeton Center for Complex Materials, a MRSEC supported by NSF Grant DMR 1420541. T.-R.C. and H.-T.J. are supported by Ministry
of Science and Technology, Academia Sinica, and National Tsing Hus University, Taiwan, and also thank NCHC, CINC-NTU, and NCTS, Taiwan, for technical support. F.S. acknowledge the National Key Projects for Basic Research of China (No:2013CB922103) and the National Natural Science Foundation of China (Grant Nos: 91421109, 11134005, 11522432). A.B. is supported by the US Department of Energy (DOE), Office of Science, Basic Energy Sciences Grant No. DE-FG02-07ER46352, and benefited from Northeastern University's Advanced Scientific Computation Center (ASCC) and the NERSC supercomputing center through DOE Grant No. DE-AC02-05CH11231.

\newpage

\clearpage
\begin{figure}
\includegraphics[width=130mm]{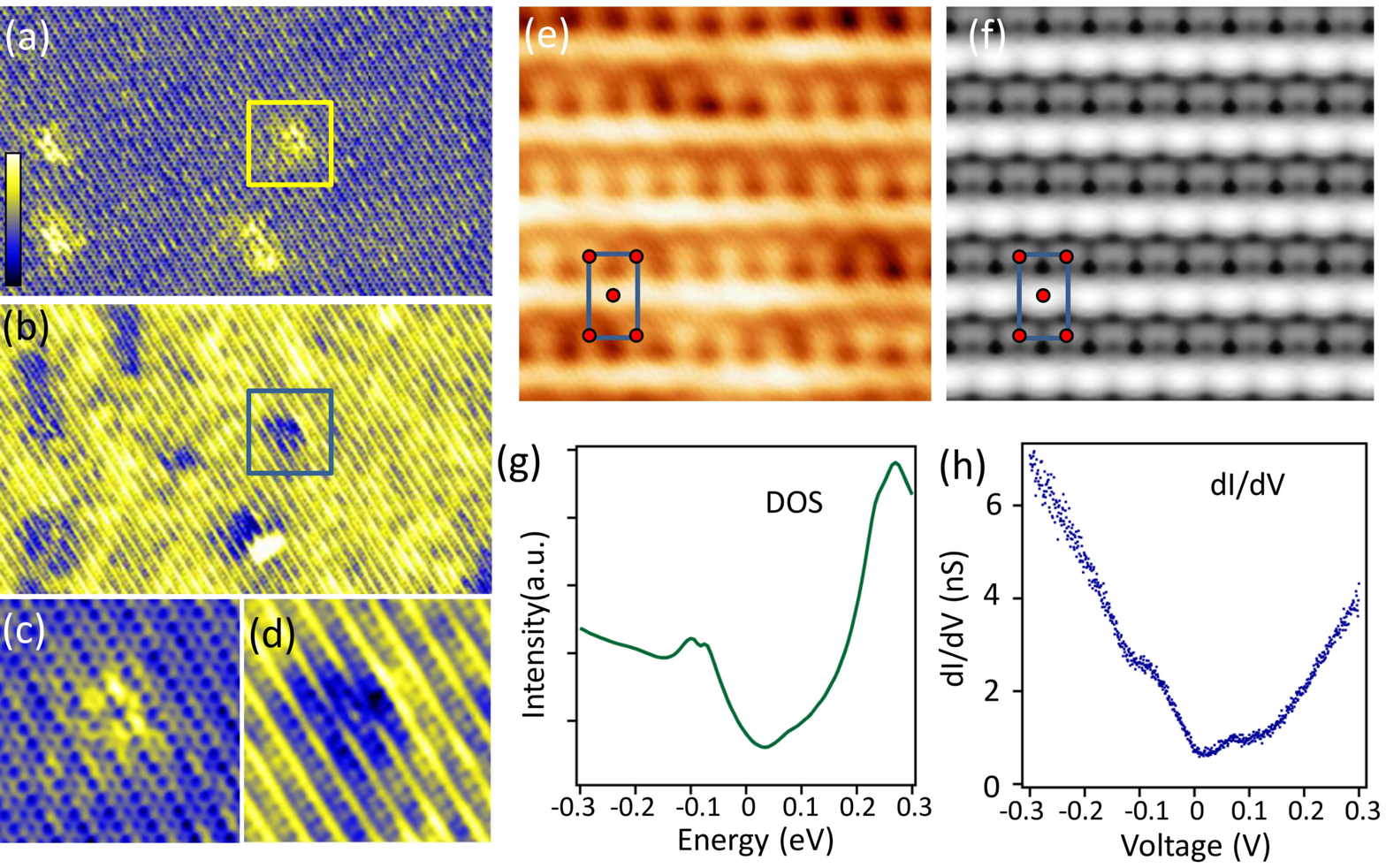}
\caption{
\label{Fig1} 
(a),(b) Large-scale constant-current STM images (30 $\times$ 18 nm$^2$) of the Mo$_{0.66}$W$_{0.34}$Te$_2$ (001) surface taken at 100~mV and -100~mV, respectively. 
White arrows indicate the crystalline orientations.
(c) and (d) are the zoom-in images (5.3 $\times$ 5.3 nm$^2$) of the defect inside the square in (a) and (b), respectively.
(e) and (f) Atomically resolved experimental and simulated STM images (-100~mV), respectively. 
In both images, brighter color means higher charge density. 
Red dots mark the positions of surface Te atoms. 
The blue rectangles indicate surface unit cells.
(g) The calculated density of state (DOS).
(h) A typical dI/dV spectrum on the Mo$_{0.66}$W$_{0.34}$Te$_2$ sample.}
\end{figure}

\begin{figure}
\includegraphics[width=130mm]{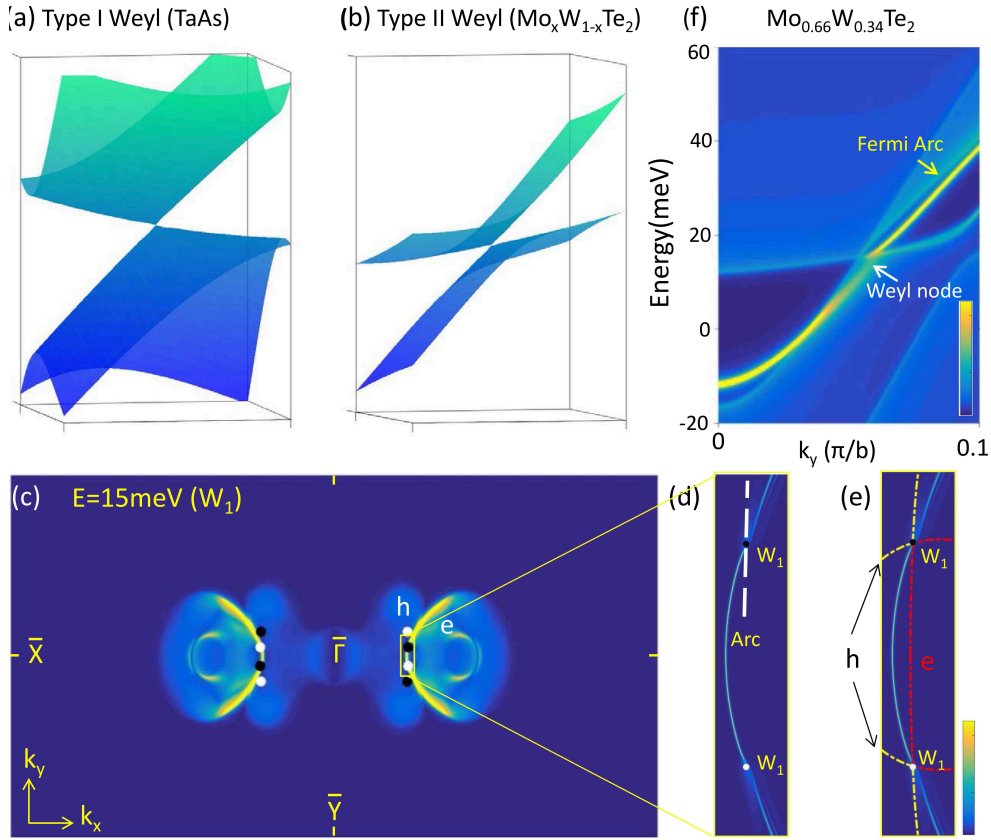}
\caption{\label{Fig2}
(a) and (b), Schematics of the type-I and the type-II Weyl cone respectively.
(c), The calculated CEC in the first surface BZ of Mo$_{0.66}$W$_{0.34}$Te$_2$ (001) at the energy of the Weyl node W$_1$. The surface weight of states is indicated by color. Projected Weyl nodes are depicted by dots. The white and black colors stand for the opposite chiralities of the Weyl nodes.
``e" (``h") stands for the electron (hole) pocket.
(d) and (e), The zoom-in views of the area inside the yellow rectangle, drawn with a lower color contrast to enhance the visibility of the surface state contours.
The topological Fermi arc, which connects one pair of projected Weyl nodes, is clearly displayed and marked.
Yellow and red dotted lines in (e) represent the boundaries of the projected hole and electron bulk pockets.
(f), The E-k dispersion cut along the white dashed line in (d).
The Weyl node and Fermi arc are marked by white and yellow arrows, respectively.
}
\end{figure}

\begin{figure*}[htbp]
	\centering
\includegraphics[width=150mm]{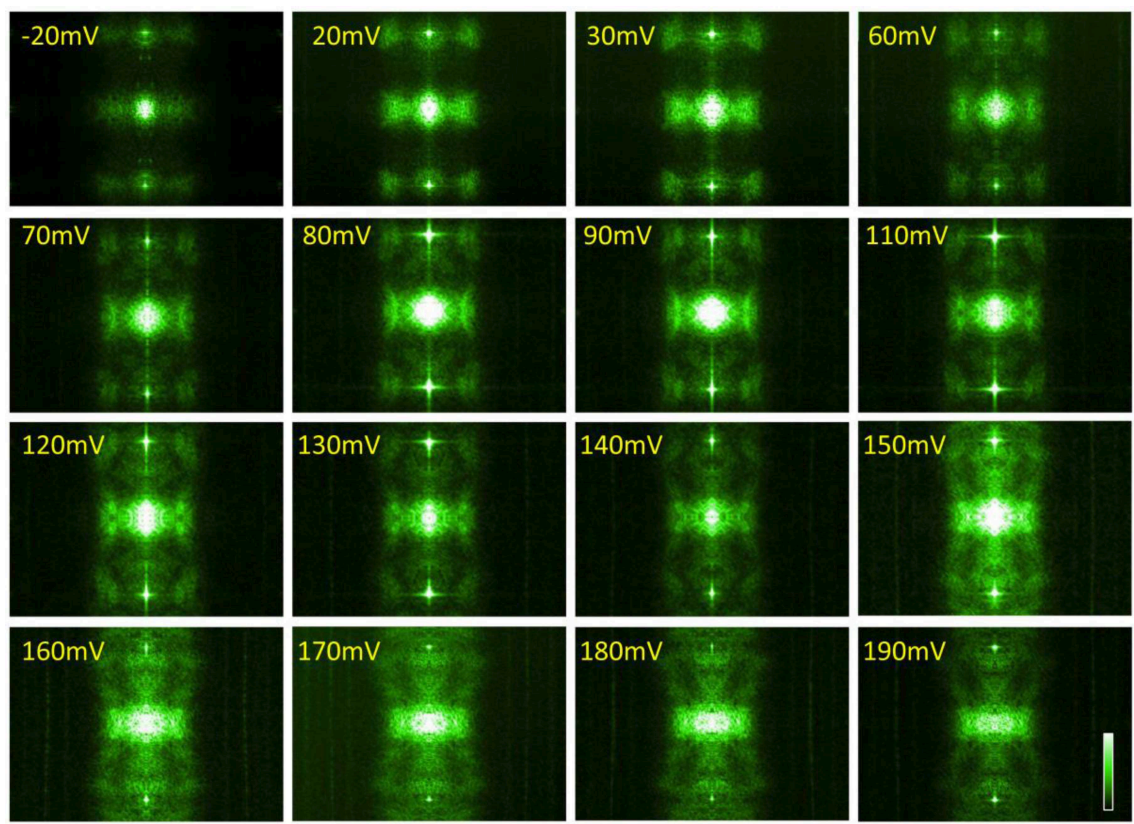}
\caption{\label{Fig3}
Energy-depdendent expeiemtnal QPI patterns measured at indicatted STM bias voltages.
}
\end{figure*}

\begin{figure*}[htbp]
	\centering
\includegraphics[width=130mm]{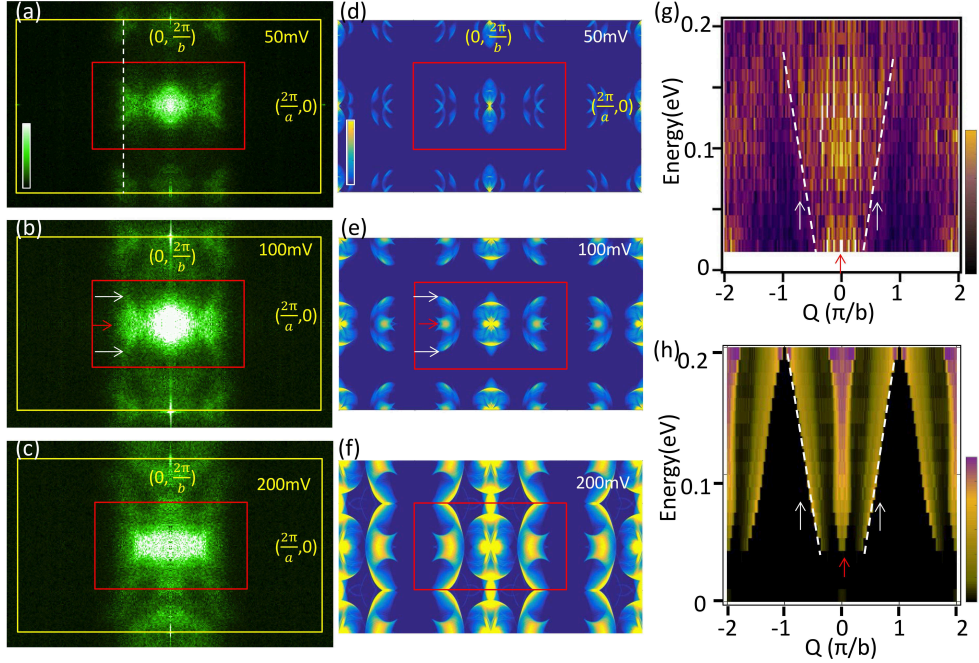}
\caption{\label{Fig4}
(a)-(c) The experimental QPI maps taken at the indicated voltages.
The areas inside red rectangles contain only features from intra-BZ scatterings.
Bragg points (($\frac{2\pi}{a}$, 0) and (0, $\frac{2\pi}{b}$)) are marked on the images.
(d)-(f) Theoretical QPI patterns derived from the surface-state-based calculations, which reasonably reproduce the features shown in the experimental data.
In (b) and (e), white arrows point to the end points of the large crescent-shaped QPI contour while red arrows mark the small central features. 
(g),(h) Experimental and theoretical E-Q dispersions taken along the dashed line in (a). The white dotted lines in (h) are drawn to mark the edges of the simulated feature and are placed to (g) as a guide to the eye.
The white (red) arrow here indicates the strongly (weakly) dispersed QPI signals as marked by the corresponding white (red) arrows in (b) and (e).
}
\end{figure*}

\begin{figure}
\includegraphics[width=130mm]{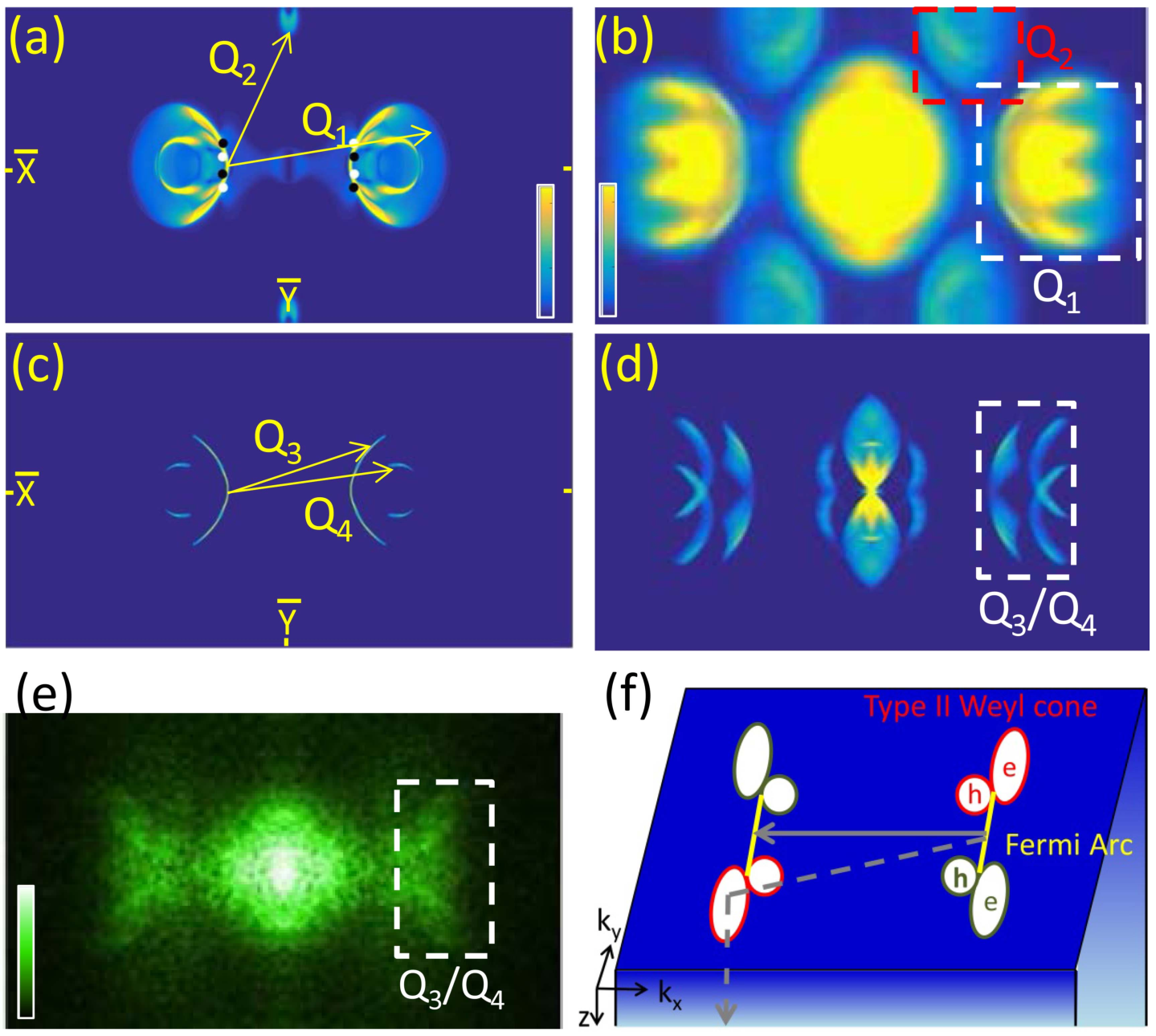}
\caption{
(a) The calculated ``complete" CEC which contains both bulk and surface state at $E$ = 50~meV.
(b) The QPI pattern based on (a), which presents only the intra-BZ scattering 
(c) The calculated CEC with only surface states considered. The Fermi arc is the central segment of the semicircular contours.
(d) The QPI pattern based on (b).
(e) The experimental QPI data (50~mV). 
The white dotted rectangles in (d) and (e) are located in the same position in Q space and are the same size.
(f) A cartoon demonstrating the sink effect of Weyl bulk states when the surface electron is scattered in to a bulk pocket, a consequence of the topological bulk-surface connection. ``e" (``h")  denotes the electron (hole) branch of a projected Weyl cone.  Straight yellow lines represent the Fermi arcs. The solid arrow indicates the scattering between two surface Fermi arc states. The dotted arrow shows the process where an electron is scattered into one branch of a Weyl cone, and sinks into the bulk.
}
\end{figure}


\begin{thebibliography}{39}

\bibitem{Weyl1} 
S. Murakami, New. J. Phys. \textbf{9}, 356 (2007).

\bibitem{Weyl2} 
X. Wan, A. M. Turner, A. Vishwanath, and S. Y. Savrasov, Phys. Rev. B \textbf{83}, 205101 (2011).

\bibitem{Weyl3}
L. Balents, Physics \textbf{4}, 36 (2011).

\bibitem{Weyl4} 
A. A. Burkov, and L. Balents, Phys. Rev. Lett. \textbf{107}, 127205 (2011).

\bibitem{Weyl5} 
T. Ojanen, Phys. Rev. B \textbf{87}, 245112 (2013).

\bibitem {rev1}
M. Z. Hasan, S.-Y. Xu, and  G. Bian, Phys. Scr. \textbf{T164}, 014001 (2015).

\bibitem {rev2}
A. M. Turner, and A. Vishwanath,	arXiv:1301.0330.

\bibitem{TaAs1}
S.-M. Huang,	S.-Y. Xu,	I. Belopolski,	C.-C. Lee,	G. Chang,	B. Wang,	N. Alidoust,	G. Bian,	M. Neupane,	C. Zhang,	S. Jia,	A. Bansil,	H. Lin,	and M. Z. Hasan, Nature Commun. \textbf{6}, 7373 (2015).

\bibitem {TaAs2}
H. Weng, C. Fang, Z. Fang, B. A. Bernevig, and X. Dai, Phys. Rev. X \textbf{5}, 011029 (2015).

\bibitem{ARPES-TaAs1}
S.-Y. Xu, I. Belopolski, N. Alidoust, M. Neupane, G. Bian, C. Zhang, R. Sankar, G. Chang, Z. Yuan, C.-C. Lee, S.-M. Huang, H. Zheng, J. Ma, D. S. Sanchez, B. Wang, A. Bansil, F. Chou, P. P. Shibayev, H. Lin S. Jia, and M. Z. Hasan,  Science \textbf{349}, 613 (2015).

\bibitem{ARPES-TaAs2}
B. Q. Lv, H. M. Weng, B. B. Fu, X. P. Wang, H. Miao, J. Ma, P. Richard, X. C. Huang, L. X. Zhao, G. F. Chen, Z. Fang, X. Dai, T. Qian, and H. Ding, Phys. Rev. X \textbf{5}, 031013 (2015).

\bibitem{CA1}
X. Huang, L. Zhao, Y. Long, P. Wang, D. Chen, Z. Yang, H. Liang, M. Xue, H. Weng, Z. Fang, X. Dai, and G. Chen, Phys. Rev. X \textbf{5}, 031023 (2015).

\bibitem{CA2}
C.-L. Zhang,	S.-Y. Xu,	I. Belopolski,	Z. Yuan,	Z. Lin,	B. Tong,	G. Bian,	N. Alidoust,	C.-C. Lee,	S.-M. Huang,	T.-R. Chang,	G. Chang,	C.-H. Hsu,	H.-T. Jeng,	M. Neupane,	D. S. Sanchez,	H. Zheng,	J. Wang,	H. Lin,	C. Zhang,	H.-Z. Lu,	S.-Q. Shen,	T. Neupert,	M. Z. Hasan	and S. Jia, Nature Commun. \textbf{7}, 10735 (2015).

\bibitem {STM-NbP}
H. Zheng, S.-Y. Xu, G. Bian, C. Guo, G. Chang, D. S. Sanchez, I. Belopolski, C.-C. Lee, S.-M. Huang, X. Zhang, R. Sankar, N. Alidoust, T.-R. Chang, F. Wu, T. Neupert, F. Chou, H.-T. Jeng, N. Yao, A. Bansil, S. Jia, H. Lin, and M. Z. Hasan, ACS Nano \textbf{10}, 1378 (2016).	
 
\bibitem {STM-TaAs1}
H. Inoue, A. Gyenis, Z. Wang, J. Li, S. W. Oh, S. Jiang, N. Ni, B. A. Bernevig, and A. Yazdani, Science \textbf{351}, 1184 (2016).

\bibitem {STM-TaAs2}
R. Batabyal, N. Morali, N. Avraham, Y. Sun, M. Schmidt, C. Felser, A. Stern, B. Yan, and H. Beidenkopf, Sci. Adv. \textbf{2}, e1600709 (2016)

\bibitem {STM-TaAs3}
P. Sessi, Y. Sun, T. Bathon, F. Glott, Z. Li, H. Chen, L. Guo, X. Chen, M. Schmidt, C. Felser, B. Yan, and M. Bode, arXiv:1606.00243.


\bibitem{WTe2}
A. A. Soluyanov,	D. Gresch,	Z. Wang,	Q. Wu,	M. Troyer,	Xi Dai, and	 B. A. Bernevig, Nature \textbf{527}, 495 (2015).


\bibitem{MoTe1}
Y. Sun, S.-C Wu, M. N. Ali, C. Felser, and B. Yan, Phys. Rev. B \textbf{92}, 161107 (2015).

\bibitem{MoTe2}
T.-R. Chang,	S.-Y. Xu,	G. Chang,	C.-C. Lee,	S.-M. Huang,	B. Wang,	G. Bian,	H. Zheng,	D. S. Sanchez,	I. Belopolski,	N. Alidoust,	M. Neupane,	A. Bansil,	H.-T. Jeng,	H. Lin, and	 M. Z. Hasan, Nature Commun. \textbf{7}, 10639 (2016).

\bibitem{MoTe3}
Z. Wang, D. Gresch, A. A. Soluyanov, W. Xie, S. Kushwaha, X. Dai, M. Troyer, R. J. Cava, and B. A. Bernevig, Phys. Rev. Lett. \textbf{117}, 056805 (2016).


\bibitem{typeII1}
A. G. Grushin, Phys. Rev. D \textbf{86}, 045001 (2012).

\bibitem{typeII2}
A. A. Zyuzin, and R. P. Tiwari, JETP Lett. \textbf{103}, 717 (2016)

\bibitem{typeII3}
H. Isobe, and N. Nagaosa, Phys. Rev. Lett. \textbf{116}, 116803 (2016)


\bibitem {STM1}
P. Roushan, J. Seo, C. V. Parker, Y. S. Hor, D. Hsieh, D. Qian, A. Richardella, M. Z. Hasan, R. J. Cava and A. Yazdani, Nature \textbf{460}, 1106 (2009).

\bibitem {STM2}
T. Zhang, P. Cheng, X. Chen, J.-F. Jia, X. Ma, K. He, L. Wang, H. Zhang, X. Dai, Z. Fang, X. Xie, and Q.-K. Xue, Phys. Rev. Lett.  \textbf{103}, 266803 (2009).

\bibitem {STM3}
I. Lee, C. K. Kim, J. Lee, S. J. L. Billinge, R. Zhong, J. A. Schneeloch, T. Liu, T. Valla, J. M. Tranquada, G. Gu, and J. C. S. Davis, Proc. Natl. Acad. Sci. U.S.A. $\mathbf{112}$, 1316 (2015).

\bibitem {STM4}
I. Zeljkovic, Y. Okada, C.-Y. Huang, R. Sankar, D. Walkup, W. Zhou, M. Serbyn, F. Chou, W.-F. Tsai, H. Lin, A. Bansil. L. Fu, M. Z. Hasan and V. Madhavan, Nat. Phys. $\mathbf{10}$, 572 (2014).

\bibitem {STM5}
D. Zhang, H. Baek, J. Ha, T. Zhang, J. Wyrick, A. V. Davydov, Y. Kuk, and J. A. Stroscio, Phys. Rev. B \textbf{89}, 245445 (2014).

\bibitem {arc1}
A. C. Potter, I. Kimchi, and A. Vishwanath, Nature Commun. \textbf{5}, 5161 (2014).

\bibitem{arc2} 
P. Hosur, Phys. Rev. B \textbf{86}, 195102 (2012).

\bibitem {QPI1}
G. Chang, S.-Y. Xu, H. Zheng, C.-C. Lee, S.-M. Huang, I. Belopolski, D. S. Sanchez, G. Bian, N. Alidoust, T.-R. Chang, C.-H. Hsu, H.-T. Jeng, A. Bansil, H. Lin, and M. Z. Hasan, Phys. Rev. Lett. \textbf{116}, 066601 (2016).

\bibitem {QPI2}
A. K. Mitchell, and L. Fritz, Phys. Rev. B \textbf{93}, 035137 (2016).

\bibitem {lattice}
B. Brown, Acta. Cryst. \textbf{20}, 268 (1966).

\bibitem {ref}
To simulate a QPI pattern in Q-space, it is a common practice to calculate the joint density of states with restrictions from spin and/or orbital textures. The simulation is simplified by considering the point defect as a $\delta$-function, by ignoring the shape of its wavefunction and scattering anisotropy of the defect. Though it may cause some minor discrepancy in the fine structures, it is widely accepted that the simulation is capable of capturing all dominated QPI features (the number, positions and sizes of QPI pockets). 

\bibitem{MoTe2-ARPES1}
I. Belopolski, S.-Y. Xu, Y. Ishida, X. Pan, P. Yu, D. S Sanchez, H.Zheng, M. Neupane, N. Alidoust, G. Chang, T.-R. Chang, Y. Wu, G. Bian, S.-M. Huang, C.-C. Lee, D. Mou, L. Huang, Y. Song, B. Wang, G. Wang, Y.-W. Yeh, N. Yao, J. E. Rault, P. Le Fevre, F. Bertran, H.-T. Jeng, T. Kondo, A. Kaminski, H. Lin, Z. Liu, F. Song, S. Shin, and M. Z. Hasan, Phys. Rev. B \textbf{94}, 085127 (2016).


\bibitem{MoTe2-ARPES2}
A. Tamai, Q. S. Wu, I. Cucchi, F. Y.  Bruno, S. Ricco, T. K. Kim, M. Hoesch, C. Barreteau, E. Giannini, C. Besnard, A. A. Soluyanov, and F. Baumberger, Phys. Rev. X \textbf{6}, 031021 (2016).

\bibitem{MoTe2-ARPES3}
K. Deng, G. Wan, P. Deng, K. Zhang, S. Ding, E. Wang, M. Yan, H. Huang, H. Zhang, Z. Xu, J. Denlinger, A. Fedorov, H. Yang, W. Duan, H. Yao, Y. Wu, S. Fan, H. Zhang, X. Chen, and S. Zhou, Nat. Phys. online published. 
\end{thebibliography}
\end{document}